\documentclass[aps,pra,amsmath,amssymb,floatfix,twocolumn,amsmath,superscriptaddress,twocolumn,nofootinbib,tighten,letterpaper]{revtex4}
\usepackage{multirow}
\usepackage{bbold}
\usepackage{subfigure}
\usepackage{color}
\usepackage{mathrsfs}
\usepackage{hyperref}
\usepackage[normalem]{ulem}
\usepackage{bm}

\usepackage{amsfonts, relsize, color}
\usepackage{graphics}
\usepackage{graphicx}
\usepackage{subfigure}
\usepackage{hyperref}
\usepackage{color}

\begin{document}
\title{\bf Higher-order topological insulators in amorphous solids}

\author{Adhip Agarwala}\email{adhip.agarwala@icts.res.in}
\affiliation{International Centre for Theoretical Sciences, Tata Institute of Fundamental Research, Bengaluru 560089, India}
\affiliation{Max-Planck-Institut f\"{u}r Physik komplexer Systeme, N\"{o}thnitzer Str. 38, 01187 Dresden, Germany}

\author{Vladimir Juri\v ci\' c}\email{vladimir.juricic@nordita.org}
\affiliation{Nordita, KTH Royal Institute of Technology and Stockholm University, Roslagstullsbacken 23,  10691 Stockholm,  Sweden}

\author{Bitan Roy}\email{bitan.roy@lehigh.edu}
\affiliation{Max-Planck-Institut f\"{u}r Physik komplexer Systeme, N\"{o}thnitzer Str. 38, 01187 Dresden, Germany}
\affiliation{Department of Physics, Lehigh University, Bethlehem, Pennsylvania, 18015, USA}

\date{\today}
\begin{abstract}
We identify the possibility of realizing higher order topological (HOT) phases in noncrystalline or amorphous materials. Starting from two and three dimensional crystalline HOT insulators, accommodating topological corner states, we gradually enhance structural randomness in the system. Within a parameter regime, as long as amorphousness is confined by outer crystalline boundary, the system continues to host corner states, yielding amorphous HOT insulators. However, as structural disorder percolates to the edges, corner states start to dissolve into amorphous bulk, and ultimately the system becomes a trivial insulator when amorphousness plagues the entire system. These outcomes are further substantiated by computing the quadrupolar (octupolar) moment in two (three) dimensions. Therefore, HOT phases can be realized in amorphous solids, when wrapped by a thin (lithographically grown, for example) crystalline layer. Our findings suggest that crystalline topological phases can be realized even in the absence of local crystalline symmetry.       
\end{abstract}

\maketitle

\section{Introduction}

Typically $d$-dimensional topological systems support gapless boundary modes of codimension one ($d_c=1$)~\cite{hasan-kane-review2010, qi-zhang-review2011, chiu-review2016}. For example, two-dimensional quantum Hall and spin Hall insulators support one-dimensional edge modes, a three-dimensional topological insulator accommodates massless Dirac fermions on two-dimensional surfaces~\cite{kane-mele2005, bernevig2006, fu-kane2006, fu-kane2007}. A similar bulk-boundary correspondence is also operative for the gapless systems, such as Dirac and Weyl semimetals~\cite{armitage-review2018}. Recently, a higher order generalization of topological systems has been proposed~\cite{benalcazar2017}, which can be realized in Bi~\cite{schindler2018}, phononic~\cite{serra-garcia2018} and photonic~\cite{noh2018,peterson} systems, and electrical circuits~\cite{imhof2018}. Namely, an $n$th order topological phase supports boundary states of codimension $d_c=n$, with corner ($d_c=d$) and hinge ($d_c=d-1$) modes standing as its prime representatives~\cite{song2017, benalcazar-prb2017,langbehn2017, xu2017, xue2018, franca2018, matsugatani2018, schindler-sciadv2018, ezawa2018, khalaf2018, vanmiert2018, wang-arxiv2018, hsu2018, queiroz2018, trifunovic2019, wang1-2018, yan2018, calugaru2019, ahn2018, okuma2018, Vliu2018, rodriguez2018, ghorashi2019}. In this language the traditional topological phases are first order in nature. While the boundary modes in conventional topological materials are guaranteed by fundamental symmetries (such as time-reversal, charge-conjugation) or no symmetry at all, the existence of corner or hinge modes crucially relies on the crystalline, such as the discrete four-fold rotational ($C_4$), symmetry. 

%%%%%%%%%%%%%%%%%%%%%%%%%%%%%%%%%%%%%%%%%%%%%%%%%%%%%%%%%%%
%%%%%%%%%%%%%%%%%%%%%%%%%%%%%%%%%%%%%%%%%%%%%%%%%%%%%%%%%%%
%%%%%%%%%%%%%%%%%%%%%%%%%%%%%%%%%%%%%%%%%%%%%%%%%%%%%%%%%%%
%%%%%%%%%%%%%%%%%%%%%%%%%%%%%%%%%%%%%%%%%%%%%%%%%%%%%%%%%%%
\begin{figure}[t!]
\includegraphics[width=0.7\linewidth]{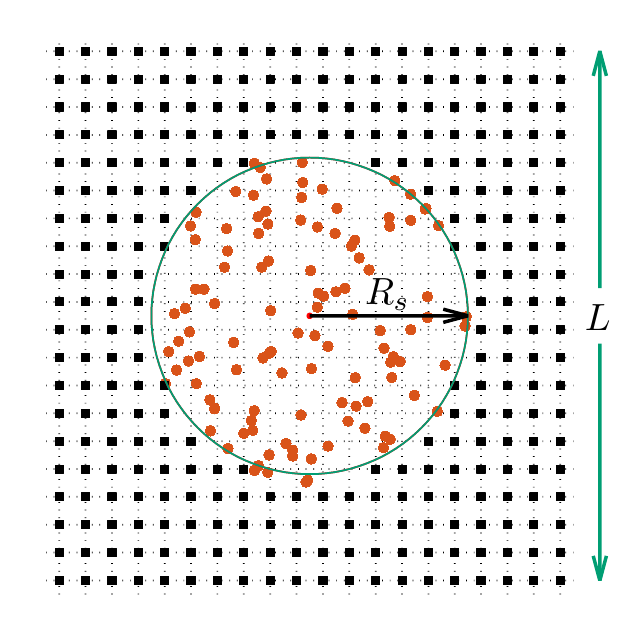}
\caption{Setup for two-dimensional amorphous HOT phases (see Fig.~\ref{Fig:AmorHOTI2D_Corner}). Here, $L$ is the linear dimension of the system. Structural disorder is confined within a radius $R_{s}$ (scrambling radius), and litters the edges when $R_s \geq L/2$. In this region, the sites are randomly distributed. This construction can be generalized to higher dimensions (see Fig.~\ref{Fig:HOTI_3D}).    
}~\label{Fig:AmorHOTI2D_Setup}
\end{figure}
%%%%%%%%%%%%%%%%%%%%%%%%%%%%%%%%%%%%%%%%%%%%%%%%%%%%%%%%%%%
%%%%%%%%%%%%%%%%%%%%%%%%%%%%%%%%%%%%%%%%%%%%%%%%%%%%%%%%%%%
%%%%%%%%%%%%%%%%%%%%%%%%%%%%%%%%%%%%%%%%%%%%%%%%%%%%%%%%%%%
%%%%%%%%%%%%%%%%%%%%%%%%%%%%%%%%%%%%%%%%%%%%%%%%%%%%%%%%%%%
%%%%%%%%%%%%%%%%%%%%%%%%%%%%%%%%%%%%%%%%%%%%%%%%%%%%%%%%%%%

Therefore, in the context of higher order topological (HOT) phases the following set of questions arises quite naturally. (a) Can HOT phases be realized in a non-crystalline environment, such as amorphous solids? (b) If so, how robust are the corresponding boundary (hinge, corner) states? We here provide an affirmative answer to the first question by introducing the concept of amorphous higher order topological insulators (HOTIs) in two and three dimensions and systematically analyze the stability of the corner states in such a medium. Our findings suggest that crystalline topological phases can be stable even in the absence of local crystalline symmetry. Given that surface states have  recently been observed in amorphous Bi$_2$Se$_3$ (possibly a $Z_2$ topological insulator)~\cite{amorphousbi2se3}, we expect that our findings will motivate future experiments demonstrating amorphous HOTI in elemental (but amorphous) Bi~\cite{schindler2018}, for example.

%%%%%%%%%%%%%%%%%%%%%%%%%%%%%%%%%%%%%%%%%%%%%%%%%%%%%%%%%%%
%%%%%%%%%%%%%%%%%%%%%%%%%%%%%%%%%%%%%%%%%%%%%%%%%%%%%%%%%%%
%%%%%%%%%%%%%%%%%%%%%%%%%%%%%%%%%%%%%%%%%%%%%%%%%%%%%%%%%%%
%%%%%%%%%%%%%%%%%%%%%%%%%%%%%%%%%%%%%%%%%%%%%%%%%%%%%%%%%%%
%%%%%%%%%%%%%%%%%%%%%%%%%%%%%%%%%%%%%%%%%%%%%%%%%%%%%%%%%%%
\begin{figure*}[t!]
\includegraphics[width=.95\linewidth]{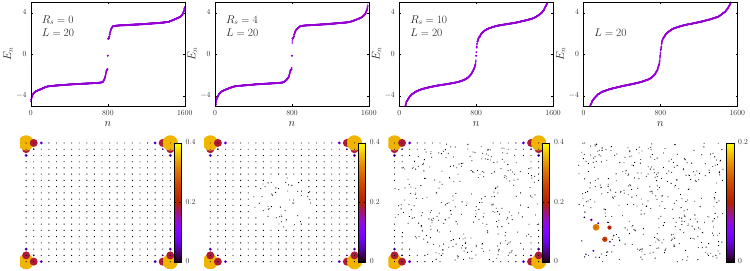}
\caption{Energy spectra (top) and the local density of states (LDOS) near (due to finite system size) zero energy modes (bottom) with increasing scrambling radius $R_s$ (from left to right) in a system with linear dimension $L=20$. Here we set $R=4$, $M=0$, $r_o=1$, $t_1=t_2=g=1$, and $n$ is the index for energy eigenvalues ($E_n$). A crystalline HOTI supports four in-gap corner modes near zero energy, well separated from the bulk states (first column). With increasing $R_s$ the bulk gap shrinks, but the system continues to support `sharp' corner states, as long as $R_s \lesssim L/2$, yielding an amorphous HOTI. When $R_s > L/2$ (entire system is scrambled), the bulk gap closes and the corner states melt into the glassy bulk (fourth column). The system then represents a trivial insulator. The scaling of the corresponding quadrupolar moment ($Q_{xy}$) with $R_s$ is shown in Fig.~\ref{Fig:AmorHOTI2D_Qxy}.
}~\label{Fig:AmorHOTI2D_Corner}
\end{figure*}
%%%%%%%%%%%%%%%%%%%%%%%%%%%%%%%%%%%%%%%%%%%%%%%%%%%%%%%%%%%
%%%%%%%%%%%%%%%%%%%%%%%%%%%%%%%%%%%%%%%%%%%%%%%%%%%%%%%%%%%
%%%%%%%%%%%%%%%%%%%%%%%%%%%%%%%%%%%%%%%%%%%%%%%%%%%%%%%%%%%
%%%%%%%%%%%%%%%%%%%%%%%%%%%%%%%%%%%%%%%%%%%%%%%%%%%%%%%%%%%
%%%%%%%%%%%%%%%%%%%%%%%%%%%%%%%%%%%%%%%%%%%%%%%%%%%%%%%%%%%

\begin{figure}[t!]
\includegraphics[width=.95\linewidth]{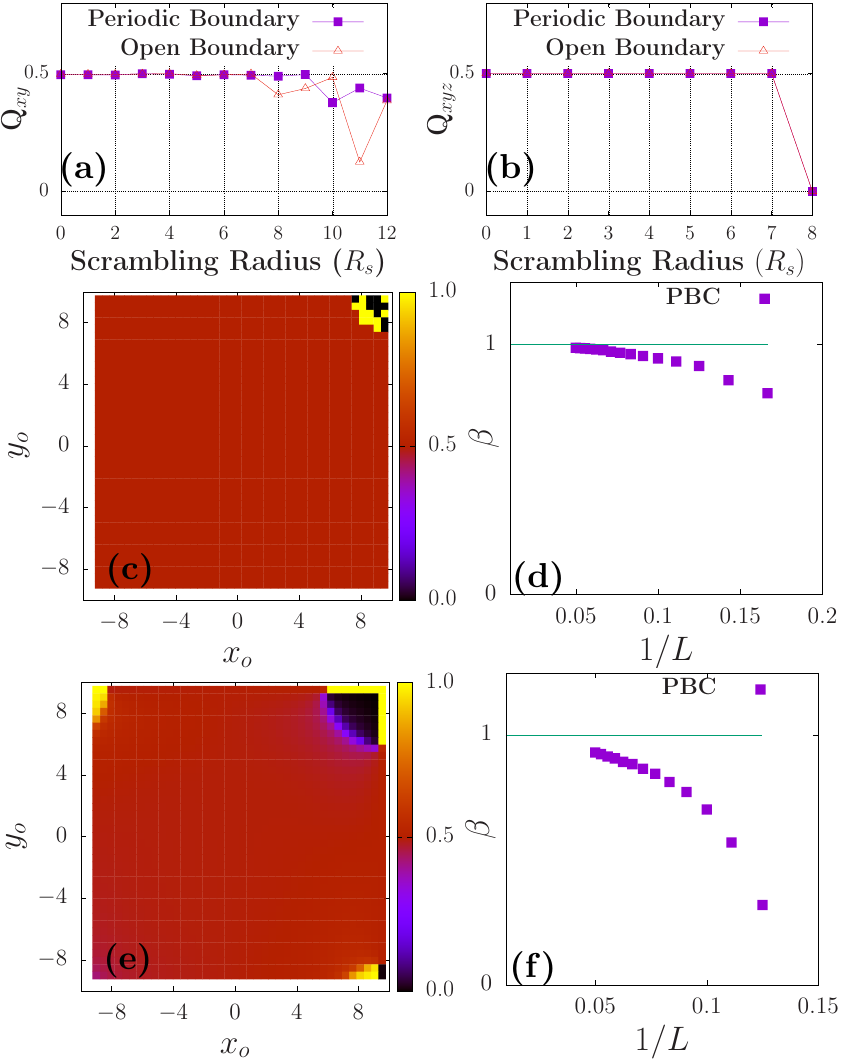}
\caption{Scaling of (a) quadrupolar ($Q_{xy}$) and (b) octupolar ($Q_{xyz}$) moments with $R_s$ for (a) $L=20$ and (b) $L=10$. In both crystalline and amorphous systems $Q_{xy}=Q_{xyz}=0.5$, when they host corner modes, see Figs.~\ref{Fig:AmorHOTI2D_Corner} and ~\ref{Fig:HOTI_3D}. For large $R_s$, the corner states disappear and $Q_{xy}$ and $Q_{xyz}$ deviate from 0.5, yielding a trivial insulator. 
Origin $(x_0,y_0)$ dependence of $Q_{xy}$ in (c) crystalline ($M=-1,g=1$) and (e) amorphous ($M=0,g=1, R_s=6,R=4, r_0=1$) HOTIs in periodic systems with $L=20$. For these sets of parameters both open crystalline and amorphous systems support four corner-localized zero-energy modes. Scaling of $\beta$, fraction of the area yielding $Q_{xy}=0.5$, with $1/L$ in periodic (d) crystalline and (f) amorphous (with $R_s=\frac{L}{2}-4$) HOTIs~\cite{Supplementary}. Therefore, minor origin dependence of quantized $Q_{xy}=0.5$ disappears in the thermodynamic limit ($L \to \infty$, with fixed $R_s/L$), promoting $Q_{xy}$ as a bulk topological invariant. A similar behavior is expected for the octupolar moment $Q_{xyz}$ in three-dimensional crystalline and amorphous HOTI.
}~\label{Fig:AmorHOTI2D_Qxy}
\end{figure}

Our main findings are the following. Starting from a two-dimensional crystalline HOTI, supporting four corner states, we systematically introduce structural disorder around its center, see Fig.~\ref{Fig:AmorHOTI2D_Setup}. When (a) amorphousness is confined within the interior of the system, and (b) the scale of structural disorder is smaller than the band gap, the corner states remain sharp and we realize a second order amorphous HOTI, see Fig.~\ref{Fig:AmorHOTI2D_Corner}. In this regime the quadrupolar moment ($Q_{xy}$) remains pinned at its quantized value of $0.5$ (within numerical accuracy), which becomes origin-independent in the thermodynamic limit (defined as $L \to \infty$, with a fixed ratio $R_s/L$), see Fig.~\ref{Fig:AmorHOTI2D_Qxy}. However, in general the corner modes start to melt in the amorphous bulk when structural disorder destroys the crystalline symmetry of the boundaries, see Fig.~\ref{Fig:AmorHOTI2D_Corner} (fourth column) and $Q_{xy}$ deviates from 0.5. Ultimately, the system becomes a trivial insulator when structural disorder plagues the entire system. Nonetheless, a significant portion of the global phase diagram of a two-dimensional amorphous insulator is occupied by HOTI, see Fig.~\ref{Fig:AmorHOTI2D_GlobalPD}. We arrive at (qualitatively) similar conclusions for a three-dimensional amorphous HOTI, see Fig.~\ref{Fig:HOTI_3D}. To summarize, it is conceivable to realize HOTIs in amorphous solids, when they are coated (grown lithographically) by a thin crystalline layer. This is in contrast to first order phases where boundary modes can be realized in completely amorphous networks~\cite{agarwala2017, mitchell2018, mansha2017, Xiao2017, Poyhonen2018, Peano2018,Chern2018}.

\section{2D Amorphous HOTI}

The Hamiltonian operator, describing a two-dimensional crystalline HOTI is 
\allowdisplaybreaks[4]
\begin{align}~\label{Eq:HOTIModel_2D}
\hat{h} &= t_1 \left[ \sigma_3 \tau_1 S_1 + \sigma_0 \tau_2 S_2 \right] + \left[ M + t_2 \left( 2-C_1 - C_2 \right) \right] \sigma_0 \tau_3 \nonumber \\
&+ g \left( C_1 -C_2\right) \sigma_1 \tau_1, 
\end{align}
where $S_j \equiv \sin (k_j a)$, $C_j \equiv \cos (k_j a)$, $k_1$ and $k_2$ are two components of spatial momenta. We set the lattice spacing $a=1$. Two sets of Pauli matrices $\{\tau_\mu\}$ and $\{ \sigma_\mu \}$ with $\mu=0, \cdots, 3$ respectively operate on the orbital and spin degrees of freedom. In real space, the above Hamiltonian operator can be equivalently characterized by a hopping matrix $ T_{j\alpha ,k\beta}(|\bf r|,\phi)$, acting between two sites $j$ and $k$ with spin/orbit indices $\alpha$ and $\beta$, respectively. Here $|\bf r|$ is the distance between two sites and the $\phi$ is the polar angle~\cite{agarwala2017, mitchell2018}. For $|{\bf r}|=0$, the hopping matrix element is described by the onsite quantity $T (0,0) = \sigma_0 \tau_3 (M+2t_2)$, while for $|{\bf r}| > 0$ it reads $T(|{\bf r}|,\phi)= \frac{t(|{\bf r}|)}{2}  \times \hat{f} (\phi)$, where
\allowdisplaybreaks[4]
\begin{equation}~\label{Eq:hoppinglongerrange}  
\hat{f} (\phi)=- i t_1\left[ \sigma_3 \tau_1  C_\phi + \sigma_0 \tau_2 S_\phi \right]  - t_2 \sigma_0 \tau_3 + g e^{i2\phi} \sigma_1 \tau_1,\end{equation}
with $t(|{\bf r}|) = \Theta(R-|{\bf r}|) \: \exp\left( -|{\bf r}|/r_o \right)$, $C_\phi = \cos(\phi)$ and $S_\phi = \sin(\phi)$. The hopping elements are bounded by a hard cut-off $R$, accompanied by an exponential decay (controlled by $r_o$), since in an amorphous solid sites are randomly distributed, see Fig.~\ref{Fig:AmorHOTI2D_Setup}. The pattern of $C_4$ symmetry breaking in Eq.~(\ref{Eq:hoppinglongerrange}) is not unique. However, the results we present here are generically insensitive to such details, as shown in the Supplemental Material~\cite{Supplementary}.

For $g=0$, a crystalline system describes a quantum spin Hall insulator, supporting two counter-propagating one dimensional edge modes for two opposite spin projections, when $0<M/t_2 <2$. The term proportional to $g$, also known as the Wilson mass, changes sign under the $C_4$ rotation and anticommutes with rest of the Hamiltonian operator. Hence, for $g \neq 0$, the boundary of the system is effectively described by one-dimensional gapless Dirac fermions in the presence of a \emph{domain wall mass}. A generalized Jackiw-Rebbi index theorem assures the presence of four corner states, where the Wilson mass changes its sign~\cite{Rackiw-Rebbi}. We then realize a second-order topological insulator. It is worth pointing out that the bulk band gap does not close (closes) across the transition between topologically distinct spin Hall (trivial) insulator and HOTI in crystalline and amorphous systems.

The Wilson mass breaks both time-reversal ($\mathcal T$) and $C_4$ symmetries, but remains invariant under a composite ${\mathcal T} C_4$ symmetry, protecting the corner modes, which get pinned at zero energy by the spectral or particle-hole symmetry generated by $\sigma_2 \tau_1$, as $\{ \hat{h}, \sigma_2 \tau_1\}=0$~\cite{half-filling-lattice}. The zero modes are also protected by an antiunitary operator $A=\sigma_3 \tau_1 K$, where $K$ is the complex conjugation, as $\{ \hat{h}, A \}=0$~\cite{BR-singleauthor}. Next we investigate the stability of such corner states in an amorphous system, lacking the $C_4$ symmetry.

%%%%%%%%%%%%%%%%%%%%%%%%%%%%%%%%%%%%%%%%%%%%%%%%%%%%%%%%%%%
%%%%%%%%%%%%%%%%%%%%%%%%%%%%%%%%%%%%%%%%%%%%%%%%%%%%%%%%%%%
%%%%%%%%%%%%%%%%%%%%%%%%%%%%%%%%%%%%%%%%%%%%%%%%%%%%%%%%%%%
%%%%%%%%%%%%%%%%%%%%%%%%%%%%%%%%%%%%%%%%%%%%%%%%%%%%%%%%%%%
%%%%%%%%%%%%%%%%%%%%%%%%%%%%%%%%%%%%%%%%%%%%%%%%%%%%%%%%%%%
%%%%%%%%%%%%%%%%%%%%%%%%%%%%%%%%%%%%%%%%%%%%%%%%%%%%%%%%%%%
%%%%%%%%%%%%%%%%%%%%%%%%%%%%%%%%%%%%%%%%%%%%%%%%%%%%%%%%%%%
%%%%%%%%%%%%%%%%%%%%%%%%%%%%%%%%%%%%%%%%%%%%%%%%%%%%%%%%%%%
%%%%%%%%%%%%%%%%%%%%%%%%%%%%%%%%%%%%%%%%%%%%%%%%%%%%%%%%%%%
%%%%%%%%%%%%%%%%%%%%%%%%%%%%%%%%%%%%%%%%%%%%%%%%%%%%%%%%%%%

\subsection{Amorphous HOTI} 

Structural disorder in a two-dimensional system of linear dimension $L$ is introduced by replacing the sites of a regular square lattice by random lattice points within a region of radius $R_s$ around the center of the system (Fig.~\ref{Fig:AmorHOTI2D_Setup}). For $R_s<L/2$, amorphousness remains confined within the interior of the system. It first reaches the boundary through the center of four edges when $R_s=L/2$, leaving the local crystalline environment around the four corners unaffected. The entire system becomes amorphous (while retaining the overall square shape) when $R_s \geq L/\sqrt{2}$. We numerically diagonalize $T (|{\bf r}|, \phi)$ for various choices of the scrambling radius $R_s$ with open boundaries and search for the localized corner states. The results are displayed in Fig.~\ref{Fig:AmorHOTI2D_Corner}.

%%%%%%%%%%%%%%%%%%%%%%%%%%%%%%%%%%%%%%%%%%%%%%%%%%%%%%%%%%%
%%%%%%%%%%%%%%%%%%%%%%%%%%%%%%%%%%%%%%%%%%%%%%%%%%%%%%%%%%%
%%%%%%%%%%%%%%%%%%%%%%%%%%%%%%%%%%%%%%%%%%%%%%%%%%%%%%%%%%%
%%%%%%%%%%%%%%%%%%%%%%%%%%%%%%%%%%%%%%%%%%%%%%%%%%%%%%%%%%%
%%%%%%%%%%%%%%%%%%%%%%%%%%%%%%%%%%%%%%%%%%%%%%%%%%%%%%%%%%%
\begin{figure}[t!]
\includegraphics[width=1.05\linewidth]{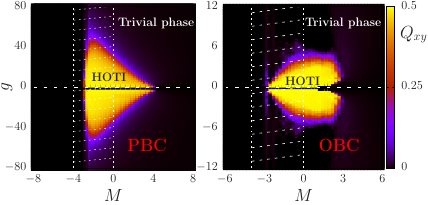}
\caption{Global phase diagram of a two-dimensional amorphous insulator in the $\left( M, g\right)$ plane [see Eq.~(\ref{Eq:HOTIModel_2D})], obtained by computing the quadrupolar moment ($Q_{xy}$) in a system with $L=20$ and $R_s=6$ with periodic (left) and open (right) boundaries. We set $R=4$ and $r_o=1$. Deep inside the amorphous HOTI (trivial phase) $Q_{xy}=0.5 (0)$. Here we average over $100$ random and independent realizations of structural disorder. In a crystalline system, the HOTI phase occupies the white dashed region. HOTI can only be found for $g \neq 0$. Along the entire $g=0$ axis, $Q_{xy}=0$ in both quantum spin Hall (for small $|M|$) and trivial (for large $|M|$) insulators, which is independent of the choice of origin for any $L$~\cite{Supplementary}.
Note that $Q_{xy}$ in periodic and open crystalline systems are identical (the white shaded region)~\cite{multipole2}. In amorphous system with PBC and OBC $Q_{xy}$ agree (within the numerical accuracy) when the scale of amorphousness or structural disorder ($|g|$) is comparable to the bandwidth ($t_1$), i.e. $|g| \lesssim 2t_1$ or much larger than the bandwidth ($|g| \gg t_1$).
}~\label{Fig:AmorHOTI2D_GlobalPD}
\end{figure}
%%%%%%%%%%%%%%%%%%%%%%%%%%%%%%%%%%%%%%%%%%%%%%%%%%%%%%%%%%%
%%%%%%%%%%%%%%%%%%%%%%%%%%%%%%%%%%%%%%%%%%%%%%%%%%%%%%%%%%%
%%%%%%%%%%%%%%%%%%%%%%%%%%%%%%%%%%%%%%%%%%%%%%%%%%%%%%%%%%%
%%%%%%%%%%%%%%%%%%%%%%%%%%%%%%%%%%%%%%%%%%%%%%%%%%%%%%%%%%%
%%%%%%%%%%%%%%%%%%%%%%%%%%%%%%%%%%%%%%%%%%%%%%%%%%%%%%%%%%%

In a crystalline system ($R_s=0$), the energy spectra display a sharp gap with precisely four corner-localized states \emph{near} zero energy, yielding a crystalline HOTI. With increasing $R_s$, the bulk gap starts to shrink, but remains finite as long as $R_s \leq L/2$, and four corner states remain sharp. The system then stands as an amorphous HOTI. Upon further increasing the scrambling radius, the corner modes start to dissolve, and ultimately disappear when $R_s \gtrsim L$, indicating onset of a trivial insulator. The corner modes start to disappear as soon as the edges loose $C_{4}$ symmetry (i.e., when $R_s \geq L/2$), although four corners still remain buried in the local crystalline environment. Therefore, amorphous HOTI is a stable phase of matter at least when the boundary of the system possesses discrete rotational symmetry, despite the bulk being plagued by structural disorder. This is the central result of our work. We further anchor this outcome from the scaling of the quadrupolar moment ($Q_{xy}$) with $R_s$.

\subsection{Quadrupolar moment ($Q_{xy}$)} 

A two-dimensional HOTI possesses a quantized quadrupolar moment $Q_{xy}$=1/2 (modulo 1)~\cite{multipole1, multipole2}, which we compute in an amorphous insulator in the following way. We consider an operator $\exp[i O]$, where 
\allowdisplaybreaks[4]
\begin{align}~\label{eq:Ooperator}
O = \frac{2\pi}{L^2}\sum_{i\alpha} f(x_{i\alpha},y_{i\alpha}) n_{i\alpha},
\end{align}
$i (\alpha)$ is the site (orbital/spin) index, and $n_{i\alpha}$ is the number operator. From the set of $N_o$ occupied states~\cite{half-filling-lattice}, schematically represented as
\allowdisplaybreaks[4]
\begin{align}
|j\rangle = \sum_{i\alpha}^{N} \psi_{i\alpha}|i\alpha\rangle, 
\end{align}
where $N$ is the total number of states and $j=1,\cdots, N_o$, we construct a $N\times N_o$ dimensional matrix $U$ by arranging $N_o$ eigenvectors columnwise. Subsequently, we introduce another matrix operator $W$, where  
\allowdisplaybreaks[4]
\begin{align}
W_{ i\alpha,j} = \exp \left[ {i \frac{2\pi}{L^2}f(x_{i\alpha},y_{i\alpha})} \right]  \: U_{ i\alpha,j }.
\end{align}
The quantity of our interest $n$ (modulo $1$) is given by 
\allowdisplaybreaks[4]
\begin{align}
n=-\frac{i}{2\pi} \; \text{Tr} \; \left[  \ln \left(U^\dagger W \right) \right].
\end{align}
To capture the topological content one needs to subtract its contribution (modulo 1) in the atomic limit, given by 
\allowdisplaybreaks[4]
\begin{align}
n_{\rm al}= n_f \sum_{i\alpha} \frac{1}{L^2} f(x_{i\alpha},y_{i\alpha}), 
\end{align} 
where $n_f=1/2$ is the filling in the system. We compute $Q_{xy} = n-n_{al}$ (modulo 1), by taking $f(x_{i\alpha},y_{i\alpha})=x_{i\alpha}y_{i\alpha}$, with both periodic and open boundaries, see Fig.~\ref{Fig:AmorHOTI2D_Qxy}.

As shown in Fig.~\ref{Fig:AmorHOTI2D_Qxy}, in a two-dimensional crystalline HOTI $Q_{xy}=0.5$ for any $|g|$, with periodic and open boundaries. In amorphous systems, $Q_{xy}$ remains pinned at 0.5 as long as $R_s  \leq L/2$, confirming the realization of an amorphous HOTI for sufficiently small $|g|$. In this parameter regime, an open system hosts corner-localized zero modes, and we find $Q_{xy}=0.5$ in open as well as periodic systems, compare Figs.~\ref{Fig:AmorHOTI2D_GlobalPD} left and right. Therefore, the bulk-boundary correspondence is operative even in amorphous insulators at least when the scale of amorphous randomness $|g| \lesssim 2t_1$, i.e., comparable to the bandwidth. By contrast, when $R_s \gtrsim L/2$, $Q_{xy}$ starts to deviate from 0.5, indicating a disappearance of an amorphous HOTI. For $R_s \gg L/2$, $Q_{xy}$ fluctuates wildly and the system becomes a trivial insulator. These findings are in qualitative agreement with the explicit computation of the corner states in a system with open boundaries, see Fig.~\ref{Fig:AmorHOTI2D_Corner} (for small $|g|$, but comparable to the bandwidth). Therefore, $Q_{xy}$ continues to be an indicator for HOTI even in amorphous solids, which we further exploit to construct a global phase diagram associated with the model in Eq.~(\ref{Eq:HOTIModel_2D}). We find that $Q_{xy}=0.5 (0.0)$ in crystalline and amorphous HOTI (trivial insulator) phases are insensitive to the choice of origin in the thermodynamic limit (defined as $L \to \infty$, keeping the ratio $R_s/L$ to be fixed)~\cite{Supplementary}, thus qualify as a topological order-parameter at least for the model we study here~\cite{watanabe-Qxy}. A similar scaling of $Q_{xy}$ has also been reported for Floquet HOTI~\cite{nag-juricic-roy}. We should note that quadrupolar moment ($Q_{xy}$) being a topological invariant is strictly defined in periodic systems, similar to the polarization (Bott index) for one-dimensional Su-Schrieffer-Heeger (two-dimensional Chern) insulator.

%%%%%%%%%%%%%%%%%%%%%%%%%%%%%%%%%%%%%%%%%%%%%%%%%%%%%
%%%%%%%%%%%%%%%%%%%%%%%%%%%%%%%%%%%%%%%%%%%%%%%%%%%%%
%%%%%%%%%%%%%%%%%%%%%%%%%%%%%%%%%%%%%%%%%%%%%%%%%%%%%
%%%%%%%%%%%%%%%%%%%%%%%%%%%%%%%%%%%%%%%%%%%%%%%%%%%%%
%%%%%%%%%%%%%%%%%%%%%%%%%%%%%%%%%%%%%%%%%%%%%%%%%%%%%
\begin{figure}[t!]
\includegraphics[width=.90\linewidth]{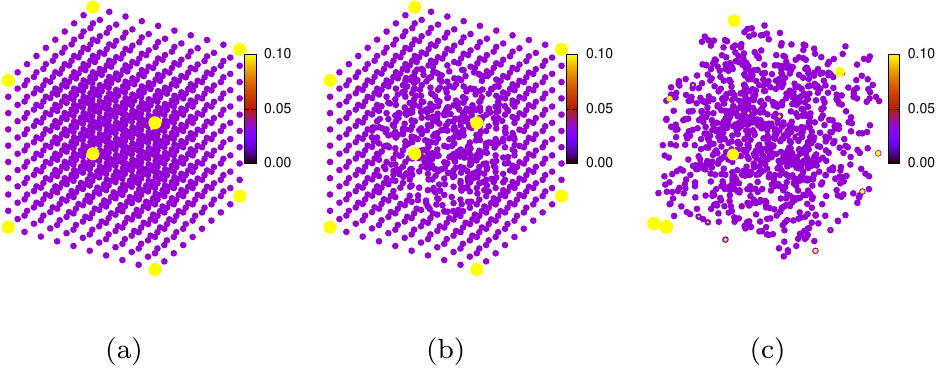}
\caption{LDOS for near zero energy states in a three dimensional cubic system with $L=10$ for (a) $R_s=0$ (crystalline HOTI), (b) $R_s=4$ (amorphous HOTI) and (c) completely scrambled (trivial insulator). We set $R=2$ and $r_0=0.5$.
}~\label{Fig:HOTI_3D}
\end{figure}
%%%%%%%%%%%%%%%%%%%%%%%%%%%%%%%%%%%%%%%%%%%%%%%%%%%%%
%%%%%%%%%%%%%%%%%%%%%%%%%%%%%%%%%%%%%%%%%%%%%%%%%%%%%
%%%%%%%%%%%%%%%%%%%%%%%%%%%%%%%%%%%%%%%%%%%%%%%%%%%%%
%%%%%%%%%%%%%%%%%%%%%%%%%%%%%%%%%%%%%%%%%%%%%%%%%%%%%
%%%%%%%%%%%%%%%%%%%%%%%%%%%%%%%%%%%%%%%%%%%%%%%%%%%%%

\subsection{Global phase diagram} 

Now we proceed to construct the global phase diagram of an amorphous insulator in the $(M,g)$ plane for a fixed scrambling radius $R_s=6$ in a system with $L=20$, see Fig.~\ref{Fig:AmorHOTI2D_GlobalPD} (left), such that structural disorder is confined within its interior. For any nontrivial $g$, there are only two possible phases: (a) amorphous HOTI and (b) trivial insulator, respectively characterized by $Q_{xy}=0.5$ and $0$ in periodic systems.

For $|M|\gg 0$, the system is always a trivial insulator for any $g$, while around $M=0$ system describes an amorphous HOTI (a trivial insulator) for small (large) $|g|$, see Fig.~\ref{Fig:AmorHOTI2D_GlobalPD} (left). This outcome is in stark contrast to that in a crystalline environment, where the system remains in the HOTI phase for arbitrary value of $g$ around $M=0$ (white shaded region in Fig.~\ref{Fig:AmorHOTI2D_GlobalPD}). This feature solely arises from the inevitable presence of structural disorder in amorphous systems that gets amplified with increasing $|g|$. As $|g|$ gets stronger, the scale of structural disorder eventually becomes comparable to the bulk band gap, indicating onset of a trivial insulator. Consequently, the parameter region over which an amorphous HOTI can be realized shrinks with increasing value of $|g|$. These universal features of the global phase diagram of an amorphous insulator are qualitatively independent of the choices of (a) the extent of the hopping, characterized by $R$ and $r_0$ and (b) the scrambling radius ($R_s$). However, with increasing $R_s$ the available space for the amorphous HOTI in the $(M,g)$ plane decreases for a fixed $L$. Notice that in the close vicinity of the phase boundaries between a trivial insulator and an amorphous HOTI, the quadrupolar moment $Q_{xy}$ vanishes smoothly. This feature is in stark contrast to the integer (half-integer) jump of the Bott index ($Q_{xy}$) across the transition between amorphous Chern (crystalline HOT) and trivial insulator phase boundaries~\cite{agarwala2017, mitchell2018}.

While $Q_{xy}$ is strictly defined in periodic system and Fig.~\ref{Fig:AmorHOTI2D_GlobalPD}(left) stands as the global phase diagram of ${\mathcal T}$ and $C_4$ symmetry breaking amorphous insulator, some qualitative connections can be made with an open system [Fig.~\ref{Fig:AmorHOTI2D_GlobalPD}(right)]. For example, irrespectively of the parameter values (such as $R_s, R, r_0$), for sufficiently small $|g|$ (but comparable to the bandwidth $\sim t_1$), $Q_{xy}=0.5$ and $0$ respectively in the amorphous HOTI and trivial insulating phase in both periodic and open systems. Moreover, in this regime an open system hosts four corner localized zero-energy modes when $Q_{xy}=0.5$, confirming the bulk-boundary correspondence. Also, for sufficiently large $|g|$ (much larger than the bandwidth) $Q_{xy}=0$ in periodic and open systems. Only for intermediate $|g|$ the $Q_{xy}$ in periodic and open systems differ. Whether this is specific to amorphous systems (littered with structural disorder) or generic to any disordered HOTI (subject to charge, bond disorder) remains to be investigated in the future. Nevertheless, we establish that HOTI supporting corner modes can be realized in amorphous systems, see Fig.~\ref{Fig:AmorHOTI2D_Corner}.

\section{3D Amorphous HOTI}

Finally, we introduce a three-dimensional amorphous HOTI, supporting eight corner modes ($d_c=3$). Once again we start from a crystalline HOTI, for which the Hamiltonian operator reads~\cite{benalcazar2017} 
\allowdisplaybreaks[4]
\begin{align}
\hat{h}_{\rm 3D}= \sum^3_{j=1} \left[ M+ t_1 S_j \right] \Gamma_j + t_2 \sum^3_{j=1} C_j \Gamma_{3+j},
\end{align}
where $\Gamma_j$s are mutually anticommuting eight dimensional Hermitian matrices~\cite{Supplementary}. We implement the above Hamiltonian operator in real space following a similar approach, introduced earlier for two dimensional systems, and set $M=0$ and $t_1=t_2=1$ for numerical analyses.

Our conclusions regarding the amorphous HOTI in three dimensions are qualitatively similar to the ones reported for its two-dimensional counterpart. Namely, an amorphous HOTI, accommodating eight corner states, is realized at least when the structural disorder is confined within the interior of the system and boundary respects the cubic symmetry, see Fig.~\ref{Fig:HOTI_3D}(b). However, the corner states start to melt in the amorphous bulk when the edges become amorphous as well, see Fig.~\ref{Fig:HOTI_3D}(c). Finally, the system becomes a trivial insulator when $R_s \sim L$.

These outcomes are substantiated from the scaling of the octupolar moment ($Q_{xyz}$) with the scrambling radius $R_s$, see Fig.~\ref{Fig:AmorHOTI2D_Qxy}(b). To compute $Q_{xyz}$, we take $O=2 \pi \sum_{i\alpha} f(x_{i \alpha},y_{i \alpha}, z_{i \alpha}) n_{i \alpha}/L^3$ [see Eq.~(\ref{eq:Ooperator})], where $f(x,y, z)=x y z$. In a crystalline HOTI $Q_{xyz}=1/2$~\cite{benalcazar2017, multipole1, multipole2}. With increasing $R_s$, $Q_{xyz}$ remains pinned at $0.5$ as long as $R_s \leq L/2$, yileding a three-dimensional amorphous HOTI. However, for $R_s \geq L/2$, the octupolar moment deviates from its quantized value, and the system becomes a trivial insulator. These features, at least for chosen set of parameter values, are insensitive to boundary conditions. We expect that a possible origin dependence of $Q_{xyz}$ is similar to that for the quadrupolar moment ($Q_{xy}$), shown in Fig.~\ref{Fig:AmorHOTI2D_Qxy}(c)-~\ref{Fig:AmorHOTI2D_Qxy}(f).

\section{Discussion and Outlook} 

To summarize, we here establish that HOT phases can be realized even when the crystalline symmetry is absent in the bulk of the system. In particular, we demonstrate that two- and three-dimensional HOTIs, supporting topological corner states, are stable in amorphous solids as long as they are coated by, for example, lithographically grown crystalline material, such that boundary of the system is devoid of any structural disorder. Our results should also be applicable (at least qualitatively) for HOTIs, supporting hinge modes, as well as HOT semimetals and superconductors. We hope that the present work will motivate furture works toward systematic exploration of amorphous HOT phases both theoretically and experimentally. We anticipate that our conclusions regarding the fate of crystalline topological phases in amorphous systems are applicable beyond the paradigm of HOTI. We expect that a recent observation of topological surface states in amorphous Bi$_2$Se$_3$~\cite{amorphousbi2se3}, should motivate future pursuit of amorphous HOTI in elemental Bi~\cite{schindler2018}, and Luttinger materials~\cite{szabo-moessner-roy}.

\acknowledgements 

A.A. acknowledges many enlightening discussions and related collaborations with Vijay B. Shenoy and Subhro Bhattacharjee. A.A. thanks Gil Young Cho for discussions. A.A. acknowledges MPG for funding through the Max Planck Partner group on strongly correlated systems at ICTS. The numerical calculations were done on the cluster {\it Tetris} at ICTS. V.J. acknowledges the support of the Swedish Research Council (Grant No. VR 2019-04735). B.R. was partially supported by the Startup grant from Lehigh University.


\begin{thebibliography}{}
\bibitem{half-filling-lattice} In any finite open system, corner states are located at finite (but very small) energies and symmetrically placed about zero energy, as $\{ \hat{h}, \sigma_2 \tau_1\}=0$. At half-filling all states at negative (positive) energies are filled (empty).
\end{thebibliography}
\end{document}